\documentstyle[11pt,IAUS212,twoside,psfig]{article}

\def\edcomment#1{\iffalse\marginpar{\raggedright\sl#1\/}\else\relax\fi}
\reversemarginpar

\def\ii{\'{\i}}

\def\etal{et al.}

\def\teff{\ifmmode T_{\rm eff} \else $T_{\mathrm{eff}}$\fi}

\def\ltsima{$\buildrel<\over\sim$}
\def\lsim{\lower.5ex\hbox{\ltsima}}

\newcommand{\hii}{H~{\sc ii}}
\newcommand{\ha}{\ifmmode {\rm H}\alpha \else H$\alpha$\fi}
\newcommand{\hb}{\ifmmode {\rm H}\beta \else H$\beta$\fi}
\newcommand{\whb}{\ifmmode W({\rm H}\beta) \else $W({\rm H}\beta)$\fi}
\newcommand{\wwr}{\ifmmode W({\rm WR}) \else $W({\rm WR})$\fi}
\newcommand{\iwr}{\ifmmode I({\rm WR})/I(\hb) \else $I({\rm WR})/I(\hb)$\fi}

\newcommand{\Heii}{He~{\sc ii} $\lambda$4686}
\newcommand{\qh}{\ifmmode q({\rm H}) \else $q({\rm H})$\fi}
\newcommand{\qhe}{\ifmmode q({\rm He^0}) \else $q({\rm He^0})$\fi}
\newcommand{\qhep}{\ifmmode q({\rm He^+}) \else $q({\rm He^+})$\fi}
\newcommand{\Qh}{\ifmmode Q({\rm H}) \else $Q({\rm H})$\fi}
\newcommand{\Qhe}{\ifmmode Q({\rm He^0}) \else $Q({\rm He^0})$\fi}
\newcommand{\Qhep}{\ifmmode Q({\rm He^+}) \else $Q({\rm He^+})$\fi}
\newcommand{\Qhtwo}{\ifmmode Q({\rm LW}) \else $Q({\rm LW})$\fi}
\newcommand{\qrathe}{\ifmmode q({\rm He^0})/q({\rm H}) \else $q({\rm He^0})/q({\rm H})$\fi}
\newcommand{\qrathep}{\ifmmode q({\rm He^+})/q({\rm H}) \else $q({\rm He^+})/q({\rm H})$\fi}
\newcommand{\Qrathe}{\ifmmode Q({\rm He^0})/Q({\rm H}) \else $Q({\rm He^0})/Q({\rm H})$\fi}
\newcommand{\Qrathep}{\ifmmode Q({\rm He^+})/Q({\rm H}) \else $Q({\rm He^+})/Q({\rm H})$\fi}
\newcommand{\Qhave}{\ifmmode \bar{Q}({\rm H}) \else $\bar{Q}({\rm H})$\fi}
\newcommand{\Qheave}{\ifmmode \bar{Q}({\rm He^0}) \else $\bar{Q}({\rm He^0})$\fi}
\newcommand{\Qhepave}{\ifmmode \bar{Q}({\rm He^+}) \else $\bar{Q}({\rm He^+})$\fi}
\newcommand{\Qhtwoave}{\ifmmode \bar{Q}({\rm H}_2) \else $\bar{Q}({\rm H}_2)$\fi}
\newcommand{\Qratheave}{\ifmmode \bar{Q}({\rm He^0})/\bar{Q}({\rm H}) \else $\bar{Q}({\rm He^0})/\bar{Q}({\rm H})$\fi}
\newcommand{\Qrathepave}{\ifmmode \bar{Q}({\rm He^+})/\bar{Q}({\rm H}) \else $\bar{Q}({\rm He^+})/\bar{Q}({\rm H})$\fi}

\def\micron{$\mu$m}

\def\msun{\ifmmode M_{\odot} \else M$_{\odot}$\fi}
\def\zsun{\ifmmode Z_{\odot} \else Z$_{\odot}$\fi}
\def\lsun{\ifmmode L_{\odot} \else L$_{\odot}$\fi}

\def\mup{\ifmmode M_{\rm up} \else M$_{\rm up}$\fi}
\def\mlow{\ifmmode M_{\rm low} \else M$_{\rm low}$\fi}
\newcommand{\dt}{\ifmmode \Delta t \else $\Delta t$\fi}

\def\aap{A\&A}
\def\aaps{A\&AS}

\def\apj{ApJ}
\def\apjl{ApJ}
\def\apjs{ApJS}
\def\mnras{MNRAS}

\newcommand{\oh}{\ifmmode 12 + \log({\rm O/H}) \else$12 + \log({\rm
O/H})$\fi}
\newcommand{\Ciii}{C~{\sc iii} $\lambda$5696}
\newcommand{\Civ}{C~{\sc iv} $\lambda$5808}

\begin{document}
\vspace*{1cm}
\title{The massive star IMF}
\author{Daniel Schaerer}
\affil{Laboratoire d'Astrophysique, Observatoire de
  Midi-Pyr\'en\'ees, 14 Av. E. Belin, F-31400 Toulouse, France}

\begin{abstract}
We review our current knowledge on the IMF in nearby environments,
massive star forming regions, super star clusters, starbursts and 
alike objects from studies of integrated light, 
and discuss the various techniques used to constrain the IMF.

In most cases, including UV-optical studies of stellar features
and optical-IR analysis of nebular emission,
the data is found to be compatible with a ``universal'' 
Salpeter-like IMF with a high upper mass cut-off over a large
metallicity range.
In contrast, near-IR observations of nuclear starbursts and LIRG
show indications of a lower \mup\ and/or a steeper IMF slope, for which
no alternate explanation has yet been found.
Also, dynamical mass measurements of seven super star clusters provide 
so far no simple picture of the IMF.

Finally we present recent results of a direct stellar probe of the
upper end of the IMF in metal-rich \hii\ regions, showing  
no deficiency of massive stars at high metallicity, and 
determining a lower limit on \mup\ of $\ga$ 60--90 \msun.

\end{abstract}
\section{Introduction}
\label{s_intro}
The stellar initial mass function (IMF) describes the relative
distribution of stars with different masses after their formation.
This basic quantity determines the relative radiative, chemical and 
mechanical ``production'' of stars of different masses/types.
It is is therefore of fundamental importance for a variety of issues 
in astrophysics, such as the understanding of stellar populations and
the star formation history of the Universe,
studies of the chemical evolution of galaxies, 
the interactions between stars and the interstellar medium etc.

The most fundamental question concerning the IMF is of course that of
its physical origin, which remains largely unknown today.
This issue, including e.g.\ competing theories tracing the IMF back
to fragmentation properties, negative feedback, or competitive 
accretion, is beyond the scope of the present review.
Other questions include the existence of lower and upper mass limits
of the IMF (e.g.\ does an upper mass limit \mup\ exist ? If so, is it 
``intrinsic'' -- e.g.\ due to fragmentation properties --
due to stellar self-limitation, or both ?),
and the possible dependence of the IMF on ``environmental'' conditions 
(e.g.\ metallicity, gas pressure, stellar density, background   
radiation, etc.).
Several of these questions are addressed 
in the recent proceedings edited by Gilmore \etal\ (1998).

The more ``empirical'' approach taken here is mainly to review our current 
knowledge on the IMF, its functional behaviour (single/multiple power law, 
or other), constraints on the upper and lower mass limits, 
and also techniques used to derive these quantities.
Although observations of resolved stellar populations are briefly discussed
(Sect.\ 2), the main focus of the review is on the massive star IMF in
unresolved stellar populations, including especially massive star forming
regions of various scales, i.e.\ from giant \hii\ regions to full blown 
starburst galaxies.
Another recent review on the IMF in starbursts is found in Leitherer (1998).


%

The following notation will be used subsequently.
The IMF $\chi$ is defined by $dN=\chi(m) dm$, which gives the number 
of stars with initial mass in the interval $[m,m+dm]$.
Generally the IMF is described by a powerlaw $\chi=A \, m^{-\alpha}$, 
where the Salpeter (1955) slope is given by $\alpha = 2.35$.
Other frequently used exponents are related to $\alpha$ by
$\Gamma= - x= 1 + \gamma = 1 - \alpha$.

\section{The IMF in the local Universe -- resolved populations}
Let us briefly recall what is known about the IMF from studies
of resolved stellar systems in the local Universe.

Star counts in Galactic and Magellanic Cloud (MC) 
clusters/associations reveal an
IMF with a slope close to Salpeter ($\alpha=2.3 \pm 0.3$) 
above $\sim$ 1 \msun and two turn-overs below,
as summarised in the detailed review of Kroupa (2002).
Using a correction for binaries according to Sagar \& Richtler (1991),
Kroupa (2002) lists $\alpha=2.7 \pm 0.3$. However, the reliability of
this correction method for the massive star IMF, especially for masses
$\ga$ 10 \msun, is not established. 
According to Kroupa (2002)
no statistically significant variation of the slope is found, except
seemingly for the Arches cluster analysed by Figer et al.\ (1999, cf.\ 
Figer these proceedings).

Recent studies of starburst like objects (e.g. giant \hii\ regions 
R136 in 30 Dor, NGC 3603)
have shown that low mass stars with masses down to $\sim$ 0.1--0.6 \msun\
are also formed in such environments
(e.g.\ Brandl et al. 1999, Sirianni et al. 2000).
So far, no change of the lower mass cut-off compared to more quiescent
star forming regions has thus been found.

The determination of \mup\ (if such a limit exists) 
from stellar counts requires sufficient
statistics. The minimum required stellar mass $M_{\rm tot}$ of a cluster 
in order to contain at least 1 star with mass $M_{\rm max}$
is given by $M_{\rm tot} \sim (2-3) \times 10^{3} (M_{\rm max}/100 \msun)^{1.3}$
\msun.
The range indicated here is valid for a slope $\alpha=2.3$ and \mlow=1 \msun\ 
or for the Kroupa (2002) IMF and \mlow=0.01 \msun.
E.g.\ in clusters with $M_{\rm tot} \sim 10^{4}$ \msun\ one expects thus
one star of $\sim$ 250 \msun.
Therefore a determination of the upper mass limit above $\sim$ 100 \msun\
is indeed not possible for most Galactic and MC clusters/associations 
(Massey 1998). However, in the few most massive clusters (e.g.\ Arches, R136,
Cyg OB2), provided they are also very young, this should in principle
be possible.
At the present stage, we simply note that stars with masses of the 
order of $\sim$ 100 \msun\ exist (cf.\ special session: Moffat \& Puls,
these proceedings).  

Local studies have also found no dependence of the IMF on metallicity 
(Massey et al. 1995, Massey 1998).

\section{The IMF measured in super star clusters}
\label{s_ssc}
The determination of dynamical masses of super star clusters (SSC),
based on velocity dispersion and cluster size measurements,
has been pioneered by Ho \& Filippenko (1996ab).
So far such mass determinations have been achieved for seven SSCs
(Ho \& Fillipenko 1996, Smith \& Gallagher 2001, Mengel \etal\ 2002,
Gilbert \& Graham 2001)
Together with an age determination, the comparison of their mass/light 
ratio to predictions from evolutionary synthesis models can be used
to constrain the IMF.
This technique provides likely the most direct/best available constraints 
on the IMF among the studies dealing with integrated light measurements.

The current studies suggest diverse results (see overview in Mengel
\etal\ 2002):
four SSCs tend to show L/M ratios compatible with IMF slopes close
to Salpeter, two SSCs lying in the interaction region of the
Antennae galaxies indicate a steeper IMF ($\alpha \sim 3$),
and one SSC (M82-F) shows a flatter IMF, implying a dissolution
over $\la$ 2 Gyr ( Smith \& Gallagher 2001).
No simple picture emerges from this small sample.
Given the importance of such combined dynamical and integrated light
studies, it is likely that much larger samples will be studied in
the near future.

%

\section{UV, optical, and IR studies of starbursts}
\label{s_uv_opt}

%
%

\subsection{Starbursts and high-z galaxies in the UV}
Numerous studies have analysed the rest-frame UV spectra of starbursts
from nearby objects to high redshift.
Among the stellar features detected (cf.\ review by Schaerer 2000),
the strongest ones (UV wind lines of Si~{\sc iv} $\lambda$1400, 
C~{\sc iv} $\lambda$1550, N~{\sc v} $\lambda$1240 and lines in the FUSE domain) 
can be used to constrain the parameters of the integrated  population, 
such as age, SF history, and IMF, by means of evolutionary synthesis techniques.
The most up-to-date model suited to such analysis is {\em Starburst99} 
(Leitherer \etal\ 1999, de Mello \etal\ 2000).

Summarised in one sentence ($\ldots$) the main result of these studies is that
all the objects contain young populations ($\la$ 10--20 Myr) characterised by 
continuous star formation or instantaneous bursts -- the distinction being often
difficult to draw -- which are populated with a rather normal Salpeter-like IMF
with stars up to  \mup\ $\sim$ 60 -- 100 \msun.
In a recent study Tremonti \etal\ (2001) examine
the stellar populations in the field of NGC 5253 and find a possible indication
for a steeper IMF, although other explanations (e.g.\ dissolution of clusters)
are possible.

The similarity of the spectra of many high redshift galaxies (e.g.\ Lyman break galaxies)
with local starbursts is well recognised and offers many exciting possibilities.
For example, from the beautiful spectrum of the lensed z $\sim$ 2.7 Galaxy 
MS 1512-CB58
of Pettini \etal\ (2000) these authors and de Mello \etal\ (2000) derive 
a constant star formation, and IMF slope between Salpeter and $\sim$ 2.8.
A spectral analysis of the lensed galaxy S2 by Le Borgne \etal\ (2002) also
finds compatibility with a Salpeter IMF although time dependent dust obscuration
(cf.\ Leitherer \etal\ 2002) may need to be invoked.
Analysis from overall SEDs do not provide strong constraints on the IMF
(Papovich \etal\ 2001).

\subsection{Optical studies of \hii\ galaxies and alike objects}
The optical spectra of massive star forming regions show both nebular and
stellar lines and provide thus indirect (nebular) and direct (stellar) 
information on their stellar content, and thus information on the IMF.
The former case is discussed in Sect.\ 5.

Among the stellar features detected in the optical (for a review see Schaerer
2000) are the Wolf-Rayet features (broad emission lines of \Heii, \Civ, \Ciii,
possibly also N~{\sc iii} $\lambda$4512, Si~{\sc iii} $\lambda$4565)
which are observed in objects covering a large range of metallicity
($1/50 \la Z/\zsun \ \la 2$).
Catalogues of these ``WR galaxies'' have been compiled by Conti (1991)
and Schaerer \etal\ (1999b).
Studies on WR galaxies (mostly BCD, Irr, spirals) are summarised in the 
reviews of Schaerer (1999ab). 

Including the detections of spectral signatures from both WN and WC stars
in a fair number of objects covering a large metallicity range, the following 
overall conclusions emerge from the studies of Schaerer \etal\ (1999a) and
Guseva \etal\ (2000). Except possibly at the lowest metallicities
a good agreement is found between the observations and the evolutionary synthesis
models of Schaerer \& Vacca (1998). From this comparison one finds clear
indications for short bursts ($\Delta t \le$ 2-4 Myr) in objects
with subsolar metallicity, an IMF compatible with Salpeter, and a 
large upper mass cut-off of the IMF, in agreement with several earlier 
studies (Mas-Hesse \& Kunth 1999, Schaerer 1996).
In addition, the observed WC/WN star ratios provide new constraints 
for mass loss and mixing scenarios in stellar evolution models
(Schaerer \etal\ 1999a), which should soon be confronted with 
predictions from rotation stellar models (cf. Maeder, Meynet, these
proceedings).

New results on the IMF metal-rich starbursts  have recently been obtained.
They are summarised in Sect.\ 6.
 
  
%

\subsection{Near-IR studies}
Observations at longer wavelengths are of great interest, 
as massive star formation occurs frequently in regions 
hidden behind significant amounts of dust at UV-optical wavelengths.

Pioneering work on prototypical nearby starburst galaxies such as M82
(distance 3.3 Mpc, where 1 arcsec corresponds to 15 pc)
has been undertaken by Rieke \etal\ (1980, 1993), who measured
a dynamical mass, the K-band and IR luminosities, $L_K$ and $L_{\rm IR}$, 
and the number of ionising 
photons of the nuclear region ($\sim$ 30 arcsec). 
From the relatively low $M/L_K$, and the large 2 $\mu$m and UV flux,
they concluded from synthesis modeling that an IMF favouring stars 
in the mass range 3-6 $\la M/\msun \la$ 10 over lower masses
was required.
Other indirect indications for such a so-called ``top-heavy'' IMF
have e.g.\ been found for M82 by Doane \& Matthews (1993).


Subsequent observations at high spatial resolution have been obtained
by several groups.
E.g.\ Satyapal \etal\ (1995, 1997) have obtained 1'' resolution Fabry-Perot
observations of Paschen-$\alpha$, Brackett-$\gamma$, and CO bandheads,
showing a strong spatial variation of extinction (with $A_V$ varying from $\sim$ 
2--12) and on average a smaller extinction than determined 
from the large aperture in the Rieke \etal\ studies.
From their analysis they conclude that there is no need for an IMF
differing from the Salpeter IMF.

K-band integral field spectroscopic observations of a 16x10 arcsec region
of M82 has been obtained by F\"orster-Schreiber \etal\ (2000, 2002).
A complex spatial structure including clusters of different ages
and varying spatial extinction (whose derived amount is dust 
geometry dependent) is found.
These studies illustrate the potential difficulty of 
obtaining robust conclusions from  large aperture observations 
and the use of ``global models''.


\subsection{The IMF in nuclear starbursts and LIRG}
CO absorption features, H recombination line fluxes, and near-IR
photometry have been used to study the stellar populations in 
luminous IR galaxies (LIRG, defined by $L > 10^{11} \lsun$).
Generally theses objects show a relatively weak recombination 
line spectrum and soft spectra, and
RSG features indicative of star formation over $\sim$ 10$^7$--10$^8$ yr
(Goldader \etal\ 1997).
As the apertures sample rather large regions, it is thought that
the overall activity can reasonably be represented by constant 
star formation.

%

Using a standard Salpeter IMF, evolutionary synthesis models 
have difficulties in reproducing simultaneously the low Br$\gamma$ 
equivalent width and the CO strength; fitting the latter implies
an overproduction of ionising photons.
This result is interpreted as a possible reduction of the upper
mass cut-off of the IMF (values of $30 \le \mup < 100 \msun$) 
and/or a steeper IMF slope (Goldader \etal\ 1997).
Similar observational trends and results are obtained for 
nuclear starbursts by Coziol \etal\ (2001).

If true, it is not clear what causes this deviation of the IMF from
the otherwise seemingly ``universal'' Salpeter IMF.
Is this related to a higher metallicity (cf.\ however Sect.\ 6), 
which could be expected in such evolved regions of galaxies, 
or possibly related to a high 
ISM pressure due to interactions in LIRGs? 


Potential problems affecting the analysis include
underlying populations which dilute/reduce $W({\rm Br}\gamma)$
(there are good indications for this, cf.\ Coziol \etal\ 2001),
absorption of ionising photons by dust (however, absorption of more
than 50 \% would be needed to reproduce the observed 
$L_{{\rm Br}\gamma}/L_{\rm IR}$), and
mixed stellar populations or discontinuous SF.
However, to date no alternative solution to the above near-IR analysis 
of LIRG and nuclear starbursts is known.

We note also that contradictory results were obtained from a detailed optical 
study of the massive star content of 6 metal-rich nuclear starbursts
(Schaerer \etal\ 2000).


\section{Stellar populations and the IMF from nebular studies}
\label{s_neb}
For many years optical studies have been used to reconstruct the stellar
content of giant \hii\ regions, \hii\ galaxies or alike objects
from their emission line spectra, including recombination lines and
forbidden metallic transitions (e.g.\ reviews by Stasi\'nska 1996,
Garc\ii a-Vargas 1996).

The development of extended grids of combined starburst and photoionisation
models at various metallicities has allowed to reproduce the main observed
emission line trends and the observed increase of the electron temperature
with decreasing metallicity. 
Concerning the IMF such comparisons have in particular shown a compatibility 
of the Salpeter IMF for metallicities down to $\sim$ 1/50 \zsun (the metallicity
of I Zw 18), 
and a presence of massive stars with \mup $\ga$ 80 \msun\
(Garc\ii a-Vargas et al.\ 1995, Stasinska \& Leitherer 1996).

Other advances e.g.\ in the understanding of nebular diagnostics, the
origin of the emission line sequences, 
and the presence of underlying populations have been made by such
extensive studies; also some interesting unsolved questions remain 
(cf.\ Stasi\'nska \etal\ 2001, Moy \etal\ 2001, Kewley \etal\ 2001,
Stasi\'nska \& Izotov 2002).

%

\subsection{ISO observations}
Interesting advances have been made with the advent of mid-IR 
spectroscopic observations
of starbursts and LIRGs observed with ISO in the 4-200 $\mu$m range and
apertures typically between 14-30 arcsec (SWS/LWS).
This wavelength range is in particular rich in atomic fine structure lines
originating in the ionised gas.

A case study of M82 using LWS (40-200 \micron) spectra was undertaken 
Colbert \etal\ (1999), who find
that the observed EL spectrum of M82 is compatible with an instantaneous
burst at ages $\sim$ 3--5 Myr, a Salpeter IMF, and a high upper mass
cut-off. However, as already pointed out earlier (Schaerer 2000), 
the shorter wavelength (SWS) data is clearly incompatible with
the Colbert \etal\ model predicting too hard a spectrum; 
their photoionisation model is underconstrained.

A different approach was adopted by F\"orster-Schreiber (1998) and
Thornley \etal\ (2000), who aim at an interpretation of the observed
relatively low excitation as traced by the [Ne~{\sc iii}] 15.5 /
[Ne~{\sc ii}] 12.8$\mu$m line ratio. 
In their model the ensemble of clusters / \hii\ regions and the
gas clouds in M82 are described by a single ``effective'' ionisation parameter,
whose value is adopted as typical for a sample of 27 starbursts in 
the starburst and photoionisation models of Thornley \etal\ (2000). 
From their modeling they conclude that the observations are compatible 
with a high upper mass cut-off ($M_{\rm up} \sim$ 50--100 \msun). 
However, to reproduce the relatively low average 
[Ne~{\sc iii}]/[Ne~{\sc ii}] ratio, short timescales of SF 
(attributed to possible negative feedback) are required.

%


This result could also be affected by several uncertainties.
E.g.\ how reliable/appropriate is the use of a single
mean ionisation parameter for such a diversity of objects ?
Also, metallicity variations, which -- as shown by Martin-Hernandez \etal\ (2002)
-- are known to affect the Neon line ratio, are not taken into
account. Finally, it is well possible that physically unrelated regions,
all included in the large ISO aperture,
contribute to the emission of Ne$^{2+}$ and Ne$^+$.
Future high spatial resolution observations (ground-based and with SIRTF)
should help to establish the reliability of mid-IR diagnostics and 
allow many interesting applications.


\subsection{Metal-rich \hii\ regions}
Emission lines in low excitation \hii\ regions have e.g.\ been used to 
study the properties of metal-rich regions in spiral galaxies.
From their analysis of the emission line trends, the observed 
[O~{\sc ii}]/[O~{\sc iii}] ratio, and He~{\sc i} $\lambda$5876/\hb\
Bresolin \etal\ (1999) found indications for a limitation of the
upper mass cut-off of the IMF of \mup\ $\la$ 35 \msun.

However, such indirect diagnostics depend strongly on the adopted
stellar tracks (e.g.\ outdated tracks from 1991 in the above study) 
and model atmospheres (cf.\ Schaerer 2000, and special
session summarised by Schaerer, Crowther \& Oey, these proceedings).
In fact, new sophisticated non-LTE line blanketed atmosphere
models for O and WR stars predict a softening of the radiation
field, as required to reconcile the photoionisation models and a higher 
value of \mup\ with the observations (Smith \etal, 2002, and these proceedings).
However, whether this can fully reproduce the observed line trends with 
a ``normal'' IMF at high metallicities remains to be shown.

We now discuss a {\em direct} approach to constrain the IMF in metal-rich
environments.


%

\begin{figure}[htb]
\centerline{\psfig{file=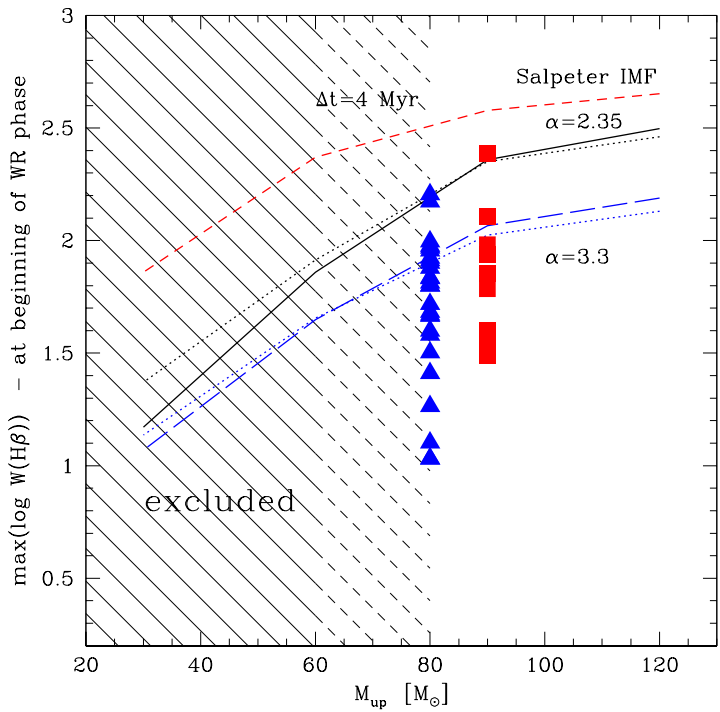,width=10cm}}
\caption{Maximum predicted \hb\ equivalent width at the beginning of 
the WR phase as a function of \mup\ for solar metallicity ($Z=0.02$)burst models 
with a Salpeter IMF (upper three curves) and a steeper IMF ($\alpha=3.3$,
lower two curves). The dotted curves show models for $Z=0.04$.
The short dashed line corresponds to an extended burst of duration $\Delta t=$ 4 Myr
(Salpeter IMF, $Z=0.02$.)
The observations are plotted at arbitrary \mup. 
The observed maximum ($\log \whb \sim$ 2.2--2.4) indicates \mup\ $\sim$ 80--90
\msun\ for a Salpeter slope, and $\protect\ga$ 120 \msun\ for $\alpha=$3.3, or somewhat
lower values for extended bursts.
}
\label{fig_imf}
\end{figure}

\section{New light on the IMF in metal-rich environments}
\label{s_metalrich}

To probe the upper end of the IMF in metal-rich environments
we have recently undertaken FORS1/VLT observations of 5 relatively 
nearby galaxies targeting known metal-rich \hii\ regions
in their disks (Pindao \etal\ 2002). 
The nuclei were avoided due to the complex mix of their stellar populations.

Spectra of $\sim$ 90 \hii\ regions with a mean metallicity close to
solar ($12+\log(O/H) \sim  8.9 \pm 0.2$) were obtained.
As suspected, we found stellar signatures of WR (WN and WC) stars in a large
number of regions, i.e.\ 27 regions plus 15 candidate WR regions.
Including previous studies (Castellanos 2001, Bresolin \& Kennicutt 2002,
Castellanos \etal\ 2002) 
our observations now nearly quadrupel the number of metal-rich \hii\ regions
where WR stars are known.
The salient result concerning the IMF is the following (see Pindao \etal\ 2002
for details).

The large sample of WR regions allows us to derive fairly model independent 
constraints on \mup\ from the maximum observed \hb\ equivalent width
of the WR regions.
Independently of the exact tracks and metallicity we derive 
a {\bf lower limit for \mup\ of 60--90 \msun} in the case of a Salpeter
slope, and larger values for steeper IMF slopes.
This constitutes a lower limit on \mup\ as all observational effects known to 
affect potentially the \hb\ equivalent width (loss of photons in slit or leakage, 
dust inside \hii\ regions, differential extinction, underlying population)
can only reduce the observed \whb. 
This result is also consistent with our previous analysis of 6 metal-rich 
nuclear starbursts with WR features indicating \mup\ $\ga$ 35-50 \msun\ 
(Schaerer \etal\ 2000).
From these analysis we conclude that the most direct stellar probes
show no deficiency of massive stars at high metallicity.

%
%

%
%
%
%

\section{Summary and conclusions}
\label{s_conclude}

In short, our current knowledge on the IMF in nearby environments,
massive star forming regions, starbursts and alike objects reviewed above 
can be summarised as follows:
\begin{itemize}
\item {\em Resolved stellar populations} show a general consensus with a 
  Salpeter-like IMF for masses $M \ga$ 1 \msun\ and the existence of stars
  with masses of $\sim$ 100 \msun\ or more.    
  Given these strong indications one may consider that the burden of proof is 
  now reversed for studies from integrated light!
\item Analysis of {\em Super Star Clusters} including dynamical mass measurements
  yield ambiguous conclusions on the slope of the IMF. Few cases are, however, 
  observed so far.
\item {\em UV-optical studies of stellar populations in HII galaxies} and alike objects
  generally show IMFs compatible with Salpeter and a large upper mass cut-off \mup.
\item {\em Near-IR studies of nuclear starburst and LIRG} show indications of a lower 
  \mup\ and/or a steeper IMF slope. No alternate explanation for the relatively 
  weak recombination line spectrum and soft spectra has so far emerged.
\item {\em Nebular studies (optical to mid-IR) of HII galaxies and starbursts}
  show no variation of the IMF from $1/50 \le Z/\zsun \la 1$.
  Indications of a lower value of \mup\ at high metallicity from nebular lines
  are probably due to inadequacies in the adopted stellar evolution / atmosphere models.
\item {\em Direct probes of WR stars in metal-rich HII regions} show  
  no deficiency of massive stars at high metallicity. 
  A lower limit on \mup\ $\ga$ 60--90 \msun\ has recently been derived from a 
  such large sample.
\end{itemize}

In most cases a seemingly ``universal'' IMF, with a powerlaw slope close to 
the Salpeter IMF is thus found. As for the physics of star formation, its origin
remains largely unknown. 

How universal is such a ``universal'' IMF ? Does this behaviour only break down
at very low metallicities, such as encountered in the earliest phases of the
Universe, and suggested by several investigations on the formation of Population
III objects (cf.\ also Abel these proceedings) ? 
These, and other challenging questions are likely to remain unanswered
for several more years.



\begin{references}
\reference Brandl, B., Brandner, 
  W., Eisenhauer, F., Moffat, A.~F.~J., Palla, F., Zinnecker, H., 1999, 
  \aap, 352, L69 
\reference Bresolin, F., Kennicutt, R. C., , 2002, \apj, 572, 838
\reference Bresolin, F., Kennicutt, R. C., Garnett, D. R., 1999, \apj, 510, 104 
\reference Castellanos, M., 2001, PhD thesis, Universida Aut\'onoma de Madrid, Spain
\reference Castellanos, M., \etal, 2002b, \mnras, submitted
\reference Colbert, J.W., et al., 1999, \apj, 511, 521
\reference Conti, P.S., 1991, \apj, 377, 115
\reference Coziol, R., Doyon, R., Demers, S., 2001, \mnras, 325, 1081 
\reference de Mello, D.F., Leitherer, C., Heckman, T.M., 2000, \apj, 530, 251 
\reference Doane, J.S., Matthews, W.G., 1993, \apj, 419, 573
\reference Figer, D.~F., Morris, M., 
  Geballe, T.~R., Rich, R.~M., Serabyn, E., McLean, I.~S., Puetter, R.~C., 
  Yahil, A., 1999, \apj, 525, 759 
\reference F\"orster-Schreiber, N.M., 1998, PhD thesis, 
    Ludwig-Maximilian-Universit\"at, Munich
\reference F\"orster-Schreiber, N.M., 2000, NewAR, 44, 263
\reference F{\"o}rster Schreiber, N.~M., Genzel, R., Lutz, D., Kunze, D., Sternberg, 
   A., 2001, \apj, 552, 544 
\reference Garc\ii a-Vargas, M.L., 1996, in ``From Stars to Galaxies'', Eds. C. Leitherer,
  U. Fritze-v. Alvensleben, J. Huchra, ASP Conf. Series, 98, 244
\reference Garc\ii a-Vargas, M.L., Bressan, A., Diaz, A., 1995, \aaps, 112, 13
\reference Gilbert, A.M., Graham, J.R., 2001, AAS, 199, 1404
\reference Gilmore, G., Howell, D. 1998, Eds., ``The Stellar Initial Mass Function'',  
  ASP Conf. Series, Vol. 142
\reference Goldader, J.D., et al., 1997, \apj, 474 104
\reference Ho, L.~C., Filippenko, A.~V., 1996, \apj, 472, 600 
\reference Ho, L.~C.,Filippenko, A.~V., 1996, \apj, 466, L83 
\reference Kewley, L.~J., Dopita, 
  M.~A., Sutherland, R.~S., Heisler, C.~A., Trevena, J., 2001, \apj, 556, 121 
\reference Kroupa, P., 2002, Science, 295, 82
\reference Le Borgne, J.F., \etal, 2002, \aap, in preparation
\reference Leitherer, C., Calzetti, D., Martins, L.~P., 2002, \apj, 574, 114 
\reference Leitherer , C., 1998, in ``The Stellar Initial Mass Function'',  
  ASP Conf. Series, Vol. 142, 61  
\reference Leitherer , C., \etal, 1999, \apjs, 123, 3 (Starburst99)
\reference Massey, P., 1998, in ``The Stellar Initial Mass Function'',  
  ASP Conf. Series, Vol. 142, 17
\reference Massey, P., Johnson, K.E., Degioia-Eastwood, K., 1995, \apj, 454, 151
\reference Mas-Hesse, J.M., Kunth, D., 1999, \aap, 349, 765
\reference Mart\ii n-Hern\'andez, N.~L., Vermeij, R., Tielens, A.~G.~G.~M., van 
der Hulst, J.~M., Peeters, E., 2002, \aap, 389, 286 
\reference Mengel, S., Lehnert, M.D., Thatte, N., Genzel, R., 2002, \aap, 383, 137
\reference Moy, E., Rocca-Volmerange, B., Fioc, M., 2001, \aap, 365, 347 
\reference Papovich, C., Dickinson, M., Ferguson, H.C., 2001, \apj, 559, 620
\reference Pettini, M., et al., 2000, \apj, 528, 96
\reference Pindao, M., Schaerer, D., Gonz\'alez Delgado, R.M., Stasi\'nska, G., 2002, \aap, 
  in press 
\reference Rieke, G.H., Lebofsky, M.J., Thompson R.I., Low, F.J., Takunaga, A.T., 
  1980, \apj, 238, 23
\reference Rieke, G.H., Loken, K., Rieke, M.J., Tamblyn, P., 1993, \apj, 412, 99
\reference Sagar, R., Richtler, T., 1991, \aap, 250, 324
\reference Salpeter, E.E, 1955, \apj, 121 161
\reference Satyapal, S. \etal, 1995, \apj, 448, 611
\reference Satyapal, S. \etal, 1997, \apj, 483, 148
\reference  Schaerer, D., 1996,\apj, 467, L17
\reference  Schaerer, D., 1999a, IAU Symp. 193, 539
\reference  Schaerer, D., 1999b,  in ``Spectrophotometric Dating of Stars and Galaxies'', 
  I. Hubeny, S.R. Heap, R.H. Cornett (eds.), ASP Conf. Series 192, 49 
\reference  Schaerer, D., 2000, in ``Stars, Gas and Dust in Galaxies: Exploring the Links'',
  Eds. D. Alloin, G. Galaz, K. Olsen, ASP Conf. Series, 221, 99
\reference  Schaerer, D., Contini, T., Kunth D., 1999a, \aap, 341, 399
\reference  Schaerer, D., Contini, T., Pindao M., 1999b, \aaps, 136, 35
\reference  Schaerer, D., Guseva, N., Izotov, Y.I., Thuan, T.X., 2000, \aap, 362, 53
\reference  Schaerer, D., Vacca, W. D. 1998, \apj, 497, 618 
\reference Sirianni, M., et al., 2000, \apj, 533, 203
\reference  Smith, L.~J., Gallagher, J.~S., 2001, \mnras, 326, 1027 
\reference  Smith L.J., Norris R.P., Crowther P.A., 2002, \mnras, in press 
  (astro-ph/0207554)
\reference Stasi\'nska, G., 1996, in ``From Stars to Galaxies'', Eds. C. Leitherer,
  U. Fritze-v. Alvensleben, J. Huchra, ASP Conf. Series, 98, 232
\reference Stasi\'nska, G., Izotov, Y.I., 2002, \aap, submitted
\reference Stasi\'nska, G., Leitherer, C., 1996, \apj, 107, 661
\reference Stasi\'nska, G., Schaerer, D., Leitherer, C., 2001, \aap, 370, 1 
\reference Thornley, M.D., Forster Schreiber, N.M., Lutz, D., 
  Genzel, R., Spoon, H. W. W., Kunze, D., Sternberg, A. 2000, \apj, 539, 641 
\reference Tremonti, C.A., Calzetti, D., Leitherer, C., Heckman, T., 2001, \apj, 555, 322
\end{references}
\end{document}